\documentstyle[aps,prl,multicol]{revtex}
\input epsf

\title{MIGHT DARK MATTER BE ACTUALLY BLACK?}

\author{Pisin Chen}
\address{
Stanford Linear Accelerator Center \\ Stanford University, Stanford, CA 94309\\
}

\begin{document}

\maketitle
\begin{abstract}
There have been proposals that primordial black hole remnants (BHRs) are the
dark matter, but the idea is somewhat vague. We argue here first that the
generalized uncertainty principle (GUP) may prevent black holes from
evaporating completely, in a similar way that the standard uncertainty
principle prevents the hydrogen atom from collapsing. Secondly we note that
the hybrid inflation model provides a plausible mechanism for production of
large numbers of small black holes. Combining these we suggest that the dark
matter might be composed of Planck-size BHRs and discuss the possible
constraints and signatures associated with this notion.
\end{abstract}

\pacs{Valid PACS appear here.
{\tt$\backslash$\string pacs\{\}} should always be input,
even if empty.}
\begin{multicols}{2}

It is by now widely accepted that dark matter (DM) constitutes a substantial
fraction of the present critical energy density in the universe. However, the
nature of DM remains an open problem. There exist many DM candidates, among
which a contending category is weakly interacting massive particles, or
WIMPs. It has been suggested that primordial black holes
(PBHs)\cite{zeldovich,hawking71} are a natural candidate for
WIMPs\cite{macgibbon}. More recent studies\cite{carr94} based on the PBH
production from the ``blue spectrum" of inflation demand that the spectral
index $n\sim 1.3$, but this possibility may be ruled out by recent CMB
observations\cite{CMB}.

In the standard view of black hole thermodynamics, based on the entropy
expression of Bekenstein\cite{bekenstein1} and the temperature expression of
Hawking\cite{hawking}, a small black hole should emit blackbody radiation,
thereby becoming lighter and hotter, leading to an explosive end when the
mass approaches zero. However Hawking's calculation assumes a classical
background metric and ignores the radiation reaction, assumptions which must
break down as the black hole becomes very small and light. Thus it does not
provide an answer as to whether a small black hole should evaporate entirely,
or leave something else behind, which we refer to as a black hole remnant
(BHR).

Numerous calculations of black hole radiation properties have been made from
different points of view\cite{wilczek}, and some hint at the existence of
remnants, but none appears to give a definitive answer. A cogent argument
against the existence of BHRs can be made\cite{susskind}: since there is no
evident symmetry or quantum number preventing it, a black hole should radiate
entirely away to photons and other ordinary stable particles and vacuum, just
like any unstable quantum system.

In a recent paper\cite{acs}, we invoked a generalized uncertainty principle
(GUP)\cite{veneziano,adler,maggiore} and argued the contrary, that the total
collapse of a black hole may be prevented by dynamics and not by symmetry,
just like the prevention of hydrogen atom from collapse by the uncertainty
principle\cite{shankar}. Our arguments then lead to a modified black hole
entropy and temperature, and as a consequence the existence of a BHR at
around the Planck mass.

In this Letter we first briefly review these arguments. We then combine this
idea with the hybrid inflation model\cite{linde91,copeland,randall,lyth99}
and show that primordial BHRs might in principle be the primary source for
dark matter.

The uncertainty principle argument for the stability of hydrogen atom can be
stated very briefly. The energy of the electron is $p^2/2m - e^2/r$, so the
classical minimum energy is very large and negative, corresponding to the
configuration $p=r=0$, which is not compatible with the uncertainty
principle. If we impose as a minimum condition that $p\approx \hbar/r$, we
see that $E=\hbar^2/2mr^2-e^2/r$, thus we find
\begin{equation}
r_{min}=\frac{\hbar^2}{me^2}\ , \quad {\rm and}\quad
E_{min}=-\frac{me^4}{2\hbar^2}\ . \label{eq:A}
\end{equation}
That is the energy has a minimum, the correct Rydberg energy, when $r$ is
the Bohr radius, so the atom is stablized by the uncertainty principle.

As a result of string theory\cite{veneziano} or general considerations of
quantum mechanics and gravity\cite{adler,maggiore}, the GUP gives the
position uncertainty as
\begin{equation}
\Delta x\geq \frac{\hbar}{\Delta p}+ l_p^2\frac{\Delta p}{\hbar} \ ,
\label{eq:B}
\end{equation}
where $l_p=(G\hbar/c^3)^{1/2}\approx 1.6\times 10^{-33}$cm is the Planck
length. A heuristic derivation may also be made on dimensional grounds. We
think of a particle such as an electron being observed by means of a photon
with momentum $p$. The usual Heisenberg argument leads to an electron
position uncertainty given by the first term in Eq.(\ref{eq:B}). But we
should add to this a term due to the gravitational interaction of the
electron with the photon, and that term must be proportional to $G$ times the
photon energy, or $Gpc$. Since the electron momentum uncertainty $\Delta p$
will be of order of $p$, we see that on dimensional grounds the extra term
must be of order $G\Delta p/c^3$, as given in Eq.(\ref{eq:B}). Note that
there is no $\hbar$ in the extra term when expressed in this way. The
position uncertainty has a minimum value of $\Delta x=2l_p$, so the Planck
distance, $l_p$, plays the role of a fundamental length.

The Hawking temperature for a spherically symmetric black hole may be
obtained in a heuristic way with the use of the standard uncertainty
principle and general properties of black holes. We picture the quantum
vacuum as a fluctuating sea of virtual particles; the virtual particles
cannot normally be observed without violating energy conservation. However
near the surface of a black hole there is an effective potential energy that
is strong enough to negate the rest energy of a particle and give it zero
total energy; of course the surface itself is a one-way membrane which can
swallow particles so that they are henceforth not observable from outside.
The net effect is that for a pair of photons one photon may be absorbed by
the black hole with effective negative energy $-E$, and the other may be
emitted to asymptotic distances with positive energy $+E$.

The characteristic energy $E$ of the emitted photons may be estimated from
the uncertainty principle. In the vicinity of the black hole surface there is
an intrinsic uncertainty in the position of any particle of about twice the
Schwarzschild radius, $\Delta x=2r_s$, due to the behavior of its field
lines\cite{adler2} - as well as on dimensional grounds. This leads to a
momentum uncertainty
\begin{equation}
\Delta p \approx \frac{\hbar}{\Delta x}=\frac{\hbar}{2r_s}=\frac{\hbar
c^2}{4GM_{\rm BH}}\ , \label{eq:C}
\end{equation}
and hence to an energy uncertainty of $\Delta pc=\hbar c^3/4GM_{\rm BH}$. We
identify this as the characteristic energy of the emitted photon, and thus as
a characteristic temperature; it agrees with the Hawking temperature up to a
factor $2\pi$, which we will henceforth include as a ``calibration factor"
and write (with $k_B=1$),
\begin{equation}
T_{\rm H} \approx \frac{\hbar c^3}{8\pi GM_{\rm BH}}=\frac{M_p^2 c^2}{8\pi
M_{\rm BH}} \ , \label{eq:D}
\end{equation}
where $M_p=(\hbar c/G)^{1/2}\approx 1.2\times 10^{19}$GeV is the Planck mass.
We know of no way to show heuristically that the emitted photons should have
a thermal blackbody spectrum except on the basis of thermodynamic
consistency.

If the energy loss is dominated by photons we may use the Stefan-Boltzmann
law to estimate the mass and energy output as functions of time. With use of
the Hawking temperature and a mass in units of the Planck mass, $x=M_{\rm
BH}/M_p$, the rate of energy loss is
\begin{equation}
\frac{dx}{dt}=\dot{x} = -\frac{1}{60(16)^2\pi t_p}\frac{1}{x^2}=
-\frac{1}{t_{ch}}\frac{1}{x^2}\ , \label{eq:E}
\end{equation}
where $t_p=(\hbar G/c^5)^{1/2}\approx 0.54\times 10^{-43}$sec is the Planck
time and $t_{ch}=60(16)^2\pi t_p\approx 4.8\times 10^4 t_p$ is a
characteristic time for BH evaporation. It follows that the mass and the
energy output rates are given by
\begin{equation}
x(t)=\Big[x_i^3-\frac{3t}{t_{ch}}\Big]^{1/3}\ ,
\end{equation}
\begin{equation}
\dot{x}=\frac{1}{t_{ch}(x_i^3-3t/t_{ch})^{2/3}}\ , \label{eq:F}
\end{equation}
where $x_i$ refers to the initial mass of the hole. The black hole
thus evaporates to zero mass in a time given by
$t/t_{ch}=x_i^3/3$, and the rate of radiation has an infinite
spike at the end of the process.

We may use the GUP to derive a modified black hole temperature.
The momentum uncertainty according to the GUP is
\begin{equation}
\frac{\Delta p}{\hbar}\approx\frac{\Delta x}{2l_p^2}\Big[1\mp
\sqrt{1-4l_p^2/(\Delta x)^2}\Big]\ . \label{eq:G}
\end{equation}
Therefore
\begin{equation}
T_{\rm GUP} = \frac{M_{\rm BH}c^2}{4\pi}\Big[1\mp\sqrt{1-1/x^2}\Big]\ .
\label{eq:H}
\end{equation}
This agrees with the Hawking result for large mass if the negative sign is
chosen, whereas the positive sign has no evident physical meaning. Note that
the temperature becomes complex and unphysical for mass less than the Planck
mass and Schwarzschild radius less than $2l_p$, the minimum size allowed by
the GUP. At the Planck mass the slope is infinite, which corresponds to zero
heat capacity of the black hole, at which the evaporation comes to a stop.

If there are $g$ species of relativistic particles, then the BH evaporation
rate is
\begin{equation}
\dot{x}=-\frac{16g}{t_{ch}}x^6\Big[1-\sqrt{1-1/x^2} \Big]^4 \ . \label{eq:I}
\end{equation}
Thus the hole with an initial mass $x_i$ evaporates to a Planck
mass remnant in a time given by
\begin{eqnarray}
\tau&=&\frac{t_{ch}}{16g}\Big[\frac{8}{3}x_i^3-8x_i-\frac{1}
{x_i}+\frac{8}{3}(x_i^2-1)^{3/2} \cr && \quad\quad\quad
-4\sqrt{x_i^2-1}+4\cos^{-1}\frac{1}{x_i} +\frac{19}{3}\Big] \cr &\approx&
\frac{x_i^3}{3g}t_{ch}, \quad\quad\quad x_i\gg 1\ . \label{eq:J}
\end{eqnarray}
The energy output given by Eq.(\ref{eq:I}) is finite at the end point where
$x=1$, i.e., $dx/dt\vert_{x=1}=-16g/t_{ch}$, whereas for the Hawking case it
is infinite at the endpoint where $x=0$. The present result thus appears to
be more physically reasonable. The evaporation time in the $x_i\gg 1$ limit
agrees with the standard Hawking picture.

The origin of BHRs is of importance and their relevance to dark matter is of
interest. Our attention is on scenarios in cosmology that can naturally
provide copious PBHs. We note that the hybrid inflation, first proposed by A.
Linde\cite{linde91}, can in principle offer that\cite{bellido}.

In the hybrid inflation model two inflaton fields, $(\phi,\psi)$, are
invoked. Governed by the inflation potential,
\begin{equation}
V(\phi,\psi)=\Big(M^2-\frac{\sqrt{\lambda}}{2}\psi^2\Big)^2
+\frac{1}{2}m^2\phi^2+\frac{1}{2}\gamma\phi^2\psi^2\ ,
\label{eq:K2}
\end{equation}
$\phi$ first executes a ``slow-roll" down the potential, and is responsible
for the more than 60 e-folds expansion while $\psi$ remains zero. When
$\phi$ eventually reduces to a critical value,
$\phi_c=(2\sqrt{\lambda}M^2/\gamma)^{1/2}$, it triggers a phase transition
that results in a ``rapid-fall" of the energy density of the $\psi$ field,
which lasts only for a few e-folds, that ends the inflation.

 The equations of motion for the fields are
\begin{eqnarray}
\ddot{\phi}+3H\dot{\phi}&=&-(m^2+\gamma\psi^2)\phi\ , \label{eq:K} \cr
\ddot{\psi}+3H\dot{\psi}&=&(2\sqrt{\lambda}M^2-\gamma\phi^2-
\lambda\psi^2)\psi\ , \label{eq:L}
\end{eqnarray}
subject to the Friedmann constraint,
\begin{equation}
H^2=\frac{8\pi}{3M_p^2}\Big[V(\phi,\psi)+\frac{1}{2}\dot{\phi}^2+
\frac{1}{2}\dot{\psi}^2\Big]\ . \label{eq:M}
\end{equation}
The solution for the $\psi$ field in the small $\phi$ regime, measured
backward from the end of inflation, is
\begin{equation}
 \psi(N(t))= \psi_e\exp(-sN(t)) \ , \label{eq:N}
\end{equation}
where $N(t)=H_*(t_e-t)$ is the number of e-folds from $t$ to $t_e$ and
$s=-3/2+(9/4+2\sqrt{\lambda}M^2/H_*^2)^{1/2}$ and
$H_*\simeq\sqrt{8\pi/3}M^2/M_p$.

We now show how large number of small black holes can result from the second
stage of inflation. Quantum fluctuations of $\psi$ induce variations of the
starting time of the second stage inflation, i.e., $\delta t =
\delta\psi/\dot{\psi}$. This translates into perturbations on the number of
e-folds, $\delta N=H_*\delta\psi/\dot{\psi}$, and therefore the curvature
contrasts.

It can be shown that\cite{liddle} the density contrast at the time when the
curvature perturbations re-enter the horizon is related to $\delta N$ by
\begin{equation}
\delta\equiv \frac{\delta\rho}{\rho}=\frac{2+2w}{5+3w}\delta N \ ,
\label{eq:O}
\end{equation}
where $p=w\rho$ is the equation of state of the universe at reentry. From
Eq.(\ref{eq:N}), it is easy to see that $\dot{\psi}=sH_*\psi$. At horizon
crossing, $\delta\psi\sim H_*/2\pi$. So with the initial condition (at
$\phi=\phi_c$) $\psi\sim H_*/2\pi$, we find that $\delta N\sim 1/s$. Thus
\begin{equation}
\delta\sim \frac{2+2w}{5+3w}\frac{1}{s}\ . \label{P}
\end{equation}
As $w$ is always of order unity, we see that the density
perturbation can be sizable if $s$ is also of order unity. With an
initial density contrast $\delta(m)\equiv \delta\rho/\rho\vert_m$,
the probability that a region of mass $m$ becomes a PBH
is\cite{carr75}
\begin{equation}
P(m) \sim \delta(m)e^{-w^2/2\delta^2} \ . \label{eq:Q}
\end{equation}

Let us assume that the universe had inflated $e^{N_c}$ times during the
second stage of inflation. From the above discussion we find\cite{bellido}
\begin{equation}
e^{N_c}\sim \Big(\frac{2M_p}{sH_*}\Big)^{1/s}\ . \label{eq:R}
\end{equation}
At the end of inflation the physical scale that left the horizon during the
phase transition is $H_*^{-1}e^{N_c}$. If the second stage of inflation is
short, i.e., $N_c\sim {\mathcal O}(1)$, then the energy soon after inflation
may still be dominated by the oscillations of $\psi$ with $p=0$, and the
scale factor of the universe after inflation would grow as $(tH_*)^{2/3}$.
The scale $(tH_*)^{2/3}H_*^{-1}e^{N_c}$ became comparable to the particle
horizon ($\sim t$), or $t\sim (tH_*)^{2/3}H_*^{-1}e^{N_c}$, when
\begin{equation}
t\sim t_h=H_*^{-1}e^{3N_c} \ . \label{eq:S}
\end{equation}
At this time if the density contrast was $\delta \sim 1$, then BHs with size
$r_s\sim H_*^{-1}e^{3N_c}$ would form with an initial mass
\begin{equation}
M_{{\rm BH}i}\simeq \frac{M_p^2}{H_*}e^{3N_c} \ . \label{eq:T}
\end{equation}

Following the numerical example given in Ref.20, and for reasons that will
become clear later, we let $H_*\sim 5\times 10^{13}$ GeV and $s\sim 3$.
Assume that the universe was radiation-dominated (so $w=1/3$) when the
curvature perturbation reentered the horizon. Then the density contrast is
$\delta \sim 1/7$, and the fraction of matter in the BH is $P(m)\sim
10^{-2}$. From Eq.(\ref{eq:R}), $e^{N_c}\sim 54$. So the total number of
e-folds is $N_c\sim 4$. The black holes were produced at the moment $t_h\sim
2\times 10^{-33}$ sec, and had a typical mass $M_{{\rm BH}i}\sim 4\times
10^{10}M_p$. Let $g\sim 100$. Then the time it took for the BHs to reduce to
remnants, according to Eq.(\ref{eq:J}), is
\begin{equation}
\tau\sim\frac{x_i^3}{3g} t_{ch}\sim 5\times 10^{-10}{\rm sec}\ . \label{eq:U}
\end{equation}
The ``black hole epoch" thus ends in time for baryogenesis and other
subsequent epochs in the standard cosmology. As suggested in Ref.20, such a
post-inflation PBH evaporation provides an interesting mechanism for
reheating. Note that due to the continuous evaporation process and the
redshift, the BH reheating should result in an effective temperature which is
sufficiently lower than the Planck scale.

This process also provides a natural way to create cold dark matter.
Although in our example $P(m)\sim 10^{-2}$, PBHs would soon dominate the
energy density by the time $t\sim P(m)^{-2}t_h\sim 2\times 10^{-29}$s,
because the original relativistic particles would be diluted much faster than
non-relativistic PBHs. By the time $t\sim\tau$, all the initial BH mass
($x_i$) had turned into radiation except one unit of $M_p$ preserved by each
BHR. As BH evaporation rate rises sharply towards the end, the universe at
$t\sim \tau$ was dominated by the BH evaporated radiation.

Roughly, $\Omega_{{\rm BHR},\tau}\sim 1/x_i$ and $\Omega_{\gamma,\tau}\sim
1$ at $t\sim \tau$, and since the universe resumed its standard evolution
after the black hole epoch ($t>\tau$), we find the density parameter for the
BHR at present to be
\begin{equation}
\Omega_{{\rm BHR},0}\sim
\Big(\frac{t_{eq}}{\tau}\Big)^{1/2}\Big(\frac{t_0}{t_{eq}}
\Big)^{2/3}\frac{1}{x_i}\Omega_{\gamma,0}\ , \label{eq:V}
\end{equation}
where $t_0\sim 4\times 10^{17}$s is the present time, and $t_{eq}$ is the
time when the density contributions from radiation and matter were equal. It
is clear from our construction that $(t_{eq}/\tau)^{1/2}\sim x_i$. So
$t_{eq}\sim 10^{12}$ sec, which is close to what the standard cosmology
assumes, and Eq.(\ref{eq:V}) is reduced to a simple and interesting
relationship:
\begin{equation}
\Omega_{{\rm BHR},0}\sim
\Big(\frac{t_0}{t_{eq}}\Big)^{2/3}\Omega_{\gamma,0}\sim
10^4\Omega_{\gamma,0}\ . \label{eq:W}
\end{equation}
In the present epoch, $\Omega_{\gamma,0}\sim 10^{-4}$. So we find
$\Omega_{{\rm BHR},0}\sim {\mathcal O}(1)$, about the right amount for dark
matter!

PBH evaporation can in principle emit late decaying massive supersymmetric
(SUSY) particles, such as gravitinos and moduli. The abundance of these
particles, and in turn that of PBHs, is limited by, e.g., big bang
nucleosynthesis (BBN), and stringent constraints have been
derived\cite{susyconstraint}. We note that these are typically based on the
gravity-mediated SUSY breaking scenario, where gravitino and moduli masses
are often at the electroweak scale and they decay at very late times. To the
contrary, in the alternative low energy gauge-mediated SUSY breaking
mechanism\cite{dimopoulos} both gravitino and moduli are light, with masses
below ${\mathcal O}(1)$eV. In addition, in this mechanism the gravitino is
naturally the lightest SUSY particle, while the decay of the ``next to
lightest SUSY particle" (NLSP) to it is very efficient. As a result of these,
in this scenario the above-mentioned constraints can be evaded.

As interactions with BHR are gravitational, the cross section is extremely
small, and direct observation of BHR seems unlikely. One possible indirect
signature may be associated with the cosmic gravitational wave background.
Unlike photons, the gravitons radiated during evaporation would be instantly
frozen. Since, according to our notion, the BH evaporation would terminate
when it reduces to a BHR, the graviton spectrum should have a cutoff at
Planck mass. Such a cutoff would have by now been redshifted to $\sim
{\mathcal O}(10^4)$ GeV. It would be interesting to investigate whether such a
spectrum is in principle observable. Another possible signature may be some
imprints on the CMB fluctuations due to the thermodynamics of BH-radiation
interactions. This will be further investigated.

I deeply appreciate my early collaborations and fruitful discussions with
Ronald J. Adler. I also thank S. Dimopoulos, A. Green, A. Linde and M.
Shmakova for helpful discussions. This work is supported by the Department of
Energy under Contract No. DE-AC03-76SF00515.

\end{multicols}

\end{document}